\documentclass[11pt]{article}
\usepackage{graphicx,xspace,oupbib,amsmath,changebar,url}

\newcommand{\ie}{\emph{i.e.}\xspace}

\newcommand{\eg}{{\em e.g.\/}\xspace}

\newcommand{\figref}[1]{\mbox{Figure~\ref{fig:#1}}}
\newcommand{\tabref}[1]{\mbox{Table~\ref{tab:#1}}}

\newcommand{\JM}[1]{{\fbox{\bf JM:} }{#1}{\fbox{\bf :end}}}
\newcommand{\PJG}[1]{{\fbox{\bf PJG:} }{#1}{\fbox{\bf :end}}}

\newcommand{\CId}[3]{#1\perp\!\!\!\perp #2\mid #3}

\newcommand{\cd}{\,|\,}
\renewcommand{\H}{\mathcal{H}}
\newcommand{\Hp}{\H_{\text{p}}}

\newcommand{\Ugt}{\textit{Ugt}}

\newcommand{\cgt}{\textit{cgt}}
\newcommand{\mgt}{\textit{mgt}}
\newcommand{\ibdyet}{\textit{ibdyet}}
\newcommand{\LR}{\text{LR}}

\newcommand{\DNAmixtures}{{\tt DNAmixtures}}
\newcommand{\RHugin}{{\tt RHugin}}
\newcommand{\Hugin}{{\tt Hugin}}
\newcommand{\R}{{\tt R}}
\newcommand{\KinMix}{{\tt KinMix}}

\begin{document}
\title{{\sc Paternity testing and other inference about relationships from DNA
mixtures}}
\author{
Peter J. Green\thanks {School of Mathematics, University of
Bristol, Bristol BS8 1TW, UK.
\newline \hspace*{5mm} Email: {\tt P.J.Green@bristol.ac.uk}.}\\
UTS, Sydney, Australia\\University of Bristol, UK.\\
\and Julia Mortera\thanks {Universit\`a Roma Tre, Italy.
\newline \hspace*{5mm} Email: {\tt julia.mortera@uniroma3.it}}\\
Universit\`a Roma Tre, Italy. }
\date{\today}
\maketitle

\begin{abstract}
We present methods for inference about relationships between contributors to a DNA mixture and other individuals of known genotype: a basic example would be testing whether a contributor to a mixture is the father of a child of known genotype. The evidence for such a relationship is evaluated as the likelihood ratio for the specified relationship versus the alternative that there is no relationship. We analyse real casework examples from a criminal case and a disputed paternity case; in both examples part of the evidence was from a DNA mixture.  DNA samples are of varying quality and therefore present challenging problems in interpretation. Our methods are based on a recent statistical model for DNA mixtures, in which a Bayesian network (BN) is used as a computational device; the present work builds on that approach, but makes more explicit use of the BN in the modelling. The R code for the analyses presented is freely available as supplementary material.  
 
We show how additional information of specific genotypes relevant to the relationship under analysis greatly strengthens the resulting inference. We find that taking full account of the uncertainty inherent in a DNA mixture can yield likelihood ratios very close to what one would obtain if we had a single source DNA profile. Furthermore, the methods can be readily extended to analyse different scenarios as our methods are not limited to the particular genotyping kits used in the examples, to the allele frequency databases used, to the numbers of contributors assumed, to the number of traces analysed simultaneously, nor to the specific hypotheses tested. 

\hspace{5mm}

\noindent {\small {\em Some key words:} Bayesian networks,
coancestry, deconvolution, disputed paternity, identity by descent, kinship, likelihood ratio.}

\end{abstract}

\section{Introduction}

This paper presents methods for inference about the relationships between contributors to a DNA mixture with unknown genotype 
and other individuals of known genotype: a basic example would be testing whether a contributor to a mixture is the father of a child of known genotype (or indeed the similar question with the roles of parent and child reversed). Following commonly accepted practice, the evidence for such a relationship is presented as the likelihood ratio for the specified relationship versus the baseline, null hypothesis, that there is no relationship at all, so the father is taken to be a random member of the population.
Our methods are based on the statistical model for DNA mixtures of \textcite{cowell:etal:15}, in which a Bayesian network (BN) is used as a computational device for efficiently computing likelihoods; the present work builds on that approach, but makes more explicit use of the BN in the modelling.

Other questions that can be answered by a similar approach include
\begin{itemize}
\item is a contributor to a mixture the brother of an individual of known genotype?
\item is a contributor to a mixture the niece of an individual of known genotype {\it and} the great-aunt of another individual of known genotype?
\item is a contributor to one mixture also a contributor to another mixture?
\item is a contributor to one mixture a brother of  a contributor to another mixture?
\item is an individual of known genotype a family relative of two contributors to a mixture who are mother and child?

\end{itemize}

A standard DNA paternity test compares the DNA profile of a putative father to that of  his alleged child; the DNA profile of the mother might or might not be available.
The case we report here (see Section \ref{sec:example}) is one of disputed inheritance. The putative father died over 20 years ago and his corpse was exhumed in order to extract his DNA profile. The
 DNA  extracted from the exhumed body sample was contaminated
and appeared to be a  mixture of at least two individuals.
 Furthermore, the DNA of the child's mother was not available.
 A preliminary analysis of this case was given in \textcite{mortera:16}. 
In that paper an approximate method based only the most probable genotype of a mixture contributor was used to specify the questioned relationship. Here we take all  uncertainty about the mixture contributors into account.

Throughout the paper, our emphasis is on methodology.
Real casework examples are presented, for illustration, but our methods are not limited to particular details of the genotyping kits, allele frequencies, number of contributors, or hypotheses in these examples.

The outline of the paper is as follows.  After a brief description of the DNA mixture model and its modification for establishing potential relationships, we introduce the  motivating example on paternity testing in Section \ref{sec:example}. Four general methods for inference about relationships from DNA mixtures are illustrated in Section \ref{sec:methods}. Results for a real case where we assess if an alleged father of a typed actor is in the mixture are given in Section \ref{sec:paternity}; results for a case where we try to identify an unknown contributor to a mixture through his potential mother's genotype are shown in Section \ref{sec:results2}. In Section \ref{sec:unions} we illustrate a proposal for computing likelihood ratios for unions of hypotheses. Indications on the available open-source software are presented in 
Section \ref{sec:software}. A general discussion and some concluding remarks are given in Section \ref{sec:discn}.

\subsection{A model for DNA mixtures}

We base the analysis of the  DNA mixture on the model described in \textcite{cowell:etal:15}. This model takes fully into account the peak heights and the possible  artefacts, like stutter and dropout, that might occur in the DNA amplification process. We give a brief summary of the main features of the model, for further details  we refer to \textcite{cowell:etal:15}. The model is an extension of the gamma model developed in  \textcite{Gammamodel} and \textcite{rgc/sll/jm:fsi}, and used in  \textcite{cowell:etal:13}. 

In summary, for a specific marker $m$ and allele $a$,  ignoring artefacts, the contribution $H_{ia}$ from an individual $i$ to the peak height at allele $a$ has a
gamma distribution, $H_{ia}\sim\Gamma(\rho\phi_in_{ia},\eta)$, where $\rho$ is proportional to the total amount of DNA in the mixture prior to
amplification; $\phi_i$ denotes the \emph{fraction} of DNA originating
from individual $i$ prior to PCR amplification, $n_{ia}$ is the number of type $a$ alleles  for individual $i$; and 
$\eta$ determines the scale. For  an amplification without artefacts of one heterozygous contributor, \ie\ $\phi_1$=1 and $n_{1,a}=1$, $\mu=\rho \eta$ is  the mean peak height and  
 $\sigma= 1/\sqrt\rho$ is the coefficient of variation. In the following we use this reparametrization. 
The model is extended to take into account artefacts: stutter, whereby a proportion of a peak belonging to allele $a$  appears as a peak at allele $a-1$; and dropout, when alleles are not observed because the peak height is below  a detection threshold $C$.  The parameter $\xi$ denotes the mean stutter proportion. 

The  evidence $E$ consists of the peak heights $\mathbf{z}$ as observed in the electropherograms,  as well
as any potential genotypes of known individuals. For given genotypes of the contributors, expressed as allele counts $\mathbf{n}=(n_{ia}, i=1,\ldots  I; a =1,\ldots, A)$, given proportions $\phi$, and given values of the parameters $(\rho,\xi,\eta)$, all observed
peak heights are independent and  for a given hypothesis $\H$,
 the full likelihood is obtained by summing over all possible
combinations of genotypes $\mathbf{n}$  with probabilities $P(\mathbf{n}\cd \H)$ associated with $\H$: 
$$
L(\H) = \Pr(E\cd \H) = \sum_{\mathbf{n}} L(\rho,\xi,\phi,\eta\cd \mathbf{z},\mathbf{n}) P(\mathbf{n}\cd \H),
$$
where \[L(\rho,\xi,\phi,\eta \cd \mathbf{z}, \mathbf{n})=\prod_m\prod_a L_{ma}(z_{ma})\] and 
\begin{equation}\label{eq:dens}
L_{ma} (z_{ma})= \left\{ \begin{array}{cr} g\{z_{ma};\rho D_a(\phi,\xi,\mathbf{n}),\eta\}& \mbox{ if $z_{ma}\geq C$}\\G\{C;\rho D_a(\phi,\xi,\mathbf{n}),\eta\}&\mbox{otherwise,}\end{array}\right.
\end{equation}
with $g$ and $G$ denoting the gamma density and cumulative distribution function respectively, and $D_a$ the effective allele counts after stutter. See \textcite{cowell:etal:15} for full details: we use their notation above.

The number of terms in this sum is huge for a hypothesis which involves several unknown contributors to the mixture, but can be calculated efficiently by Bayesian network techniques that represent the genotypes using a Markovian structure, the allele counts for each individual being modelled sequentially over the alleles. 
The maximum likelihood estimate (MLE) parameters are obtained using the \R\ package 
\DNAmixtures\ \cite{graversen:package:13}  which interfaces to the HUGIN API 
(Hugin Expert A/S, 2012) through the \R\ package \RHugin\ \cite{manual:RHugin}. 

In this paper we follow \textcite{cowell:etal:15} in estimating parameters by maximum likelihood. 
In all computations of likelihood ratios, parameters in both numerator and denominator are fixed at the MLEs under the null hypothesis.  Other choices are possible, depending on the demands of the legal environment, for example the likelihoods in the numerator and denominator could be separately maximised over values of the parameters; this would entail some additional computation.

\subsection{Relationship inference with DNA mixtures}
\label{sec:relinf}

In this work we wish to establish whether one (or more) contributors to the DNA mixture has a potential relationship with one or more individuals whose genotypes are known and who have a known relationship to each other. To do this, we make more explicit use of the BN used as a computational device in \textcite{cowell:etal:15}.

\begin{figure}[htbp]
\begin{center}
	\resizebox{0.7\textwidth}{!}{\includegraphics{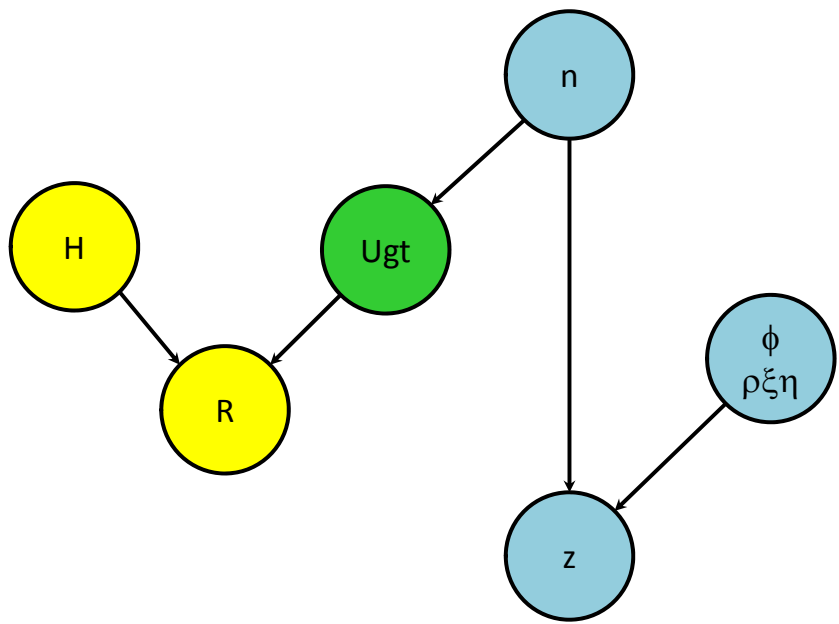}}
\end{center}
\caption{A DAG pictorial representation for establishing a potential relationship with a contributor to the mixture. The blue nodes represent the gamma model {\protect\cite{cowell:etal:15}}.
The yellow nodes denote the putative relationship between the mixture contributor $\Ugt$ and relatives with genotypes $R$, under the control of the hypothesis $\H$.}
\label{fig:DAG}
\end{figure}

This network represents the probabilistic dependence of the peak heights $\mathbf{z}$ on the allele counts $\mathbf{n}$ for the unknown contributors to the mixture, and the parameters ($\phi, \rho, \xi, \eta$) of the gamma model. This dependence is represented in the right hand part of the directed acyclic graph in \figref{DAG}.

Our general strategy is to modify the Bayes net formulation of the model of \textcite{cowell:etal:15}, in ways described in the following sections, and then, as in that earlier paper, perform the necessary computations to deliver the required likelihood ratios, as laid out by \textcite{graversen:lauritzen:comp:13}, appropriately generalised. More details on this are given in the Appendix.

\section{Motivating example: paternity testing}
\label{sec:example}
\subsection{A case study}

We now illustrate a  real case from the Forensic Institute, Sapienza Universit\`a Roma, which provides the motivating example for this paper.

\emph{A man  {$B$}, met a young lady $C$ and began a secret relationship. One of $C$'s sons {$A$}, learns as an adult 
that he is not the son of $C$'s husband but probably $B$'s son. Some years after {$B$}'s death, {$A$} claims his share of $B$'s substantial inheritance.  After his mother's death and over 20 years after {$B$}'s death, {$B$}'s body is exhumed and DNA is extracted from a bone. This is to be used to establish whether $A$ could be the son of $B$.}

This DNA is highly contaminated and appears to be {mixture of at least 2 individuals}.
\tabref{data} shows an extract of the data used for this paternity testing case.   For each marker, the first two columns of \tabref{data} show the unordered pair of alleles in the putative son $A$'s genotype. The mixed profile extracted from B's bone, is shown in columns 4 and 5,  where for each marker, we have the alleles together with their corresponding peak heights.  

This paternity testing problem offers two complicating features:   the profile from $B$ appears to be a mixture of at least two contributors and  the genotype of $A$'s mother is not available. The alleles in the mixture shared with A's genotype are italicized. 
In order to analyse this case we  need to use the  information in the peak heights.

\begin{table}[htbp]
\caption{Extract of the paternity testing data. The first two columns show the unordered pair of alleles in the child's genotype, the third column gives the markers, whereas the  alleles and  peak heights from the amplification of the bone are given in the last two columns. Italics are used to emphasise where the same allele appears in both the mixture and the son's genotype.}

\label{tab:data}
\begin{center} 
\begin{tabular}{lllcc}
\hline
\multicolumn{2}{c}{Alleged son }& & \multicolumn{2}{c}{Data from $B$'s bone} \\
\multicolumn{2}{c}{$A$'s genotype} & Marker & Alleles & {Peak height}  \\
\hline
\emph{X} & \emph{Y} &AMEL & \emph{X} & 3257   \\
 & &  & \emph{Y} & 1736    \\ 
				\hline
10 & \emph{11} & D16S539 & \emph{11} & 83  \\
   & &      & 12 & 182\\
\hline
\emph{15} & 16 & D8S1179 & 12    & 398     \\        
 & &    & 13    & 1406  \\        
 & &    & \emph{15}  & 1395 \\
\hline
\emph{30} &  32 & D21S11 & 29      & 139    \\
 & & & \emph{30} & 815    \\
   &     &  & 31      & 88     \\    
   &      &  & 31.2    & 241      \\    
   &      &  & 34      & 151      \\
\hline
\emph{13}  & 16 & D18S51 & 12 & 59      \\    
 & &  & \emph{13}  & 60     \\
\hline
\emph{14} & \emph{14} & D2S441 & \emph{14} & 3683   \\
\hline
\emph{14} & \emph{14} & D3S1358  & \emph{14} & 858  \\
 & &  & 15    & 708   \\
\hline
15.3 & \emph{17} & D1S1656 & 16 & 387 \\
 & & & \emph{17} & 326   \\
\hline
\emph{17} & 23 & D12S391 & \emph{17} & 165   \\
       & &   & 18 & 83 \\
\hline
$\cdots$ & $\cdots$ & $\cdots$ & $\cdots$ & $\cdots$ \\
\hline
14 & \emph{20} & SE33    & \emph{20} & 139  \\
\hline
\end{tabular}
 \end{center}
\end{table}

\begin{figure}[htbp]
\begin{center}
			       \resizebox{\textwidth}{!}{\includegraphics{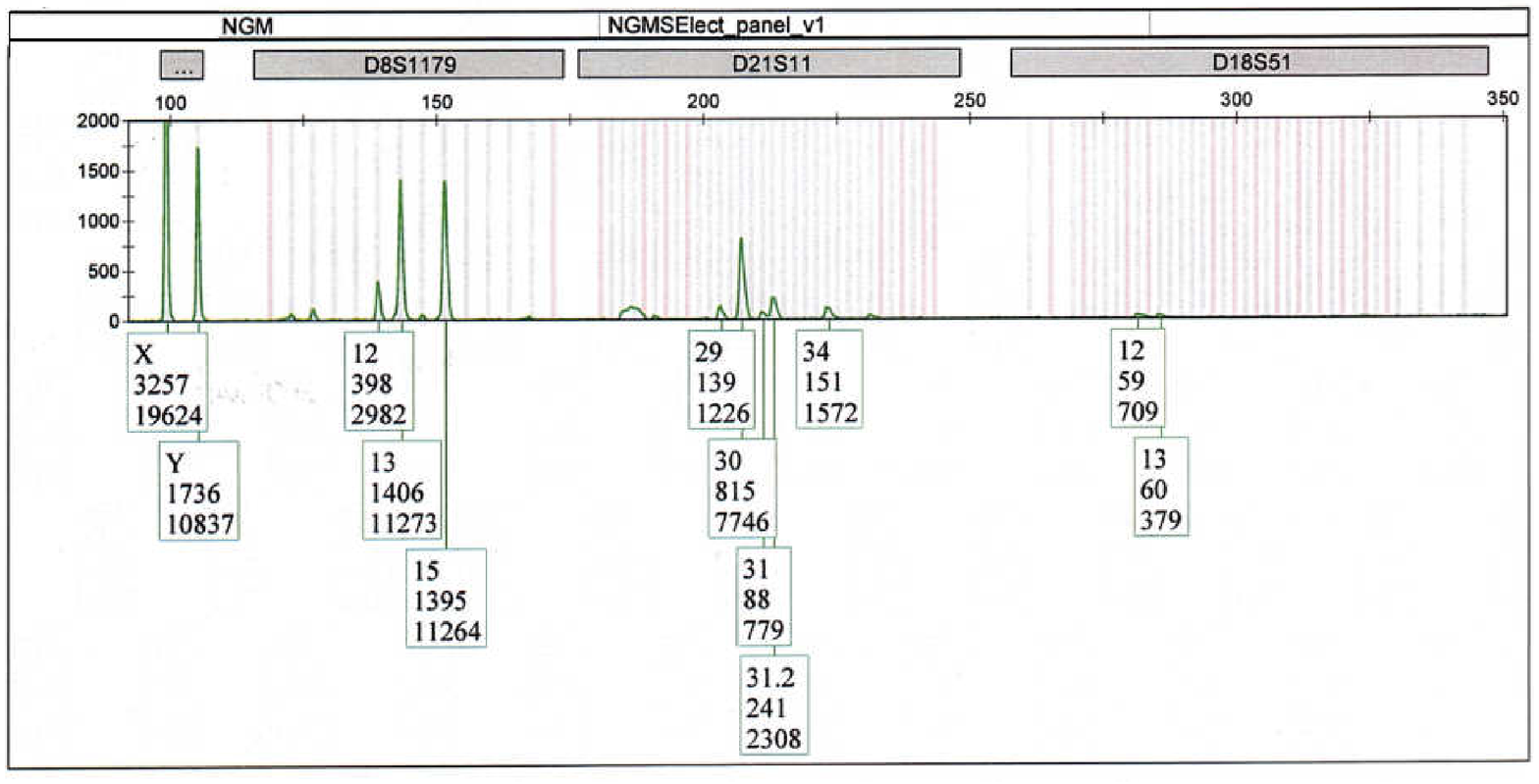}}
\end{center}
\caption{Extract from an electropherogram (EPG), showing the green dye lane.}
\label{fig:EPG}
\end{figure}

\figref{EPG} shows a portion of the original electropherogram (EPG) obtained from the bone.
There are signs that the DNA is subject to contamination, presumably due to  the fact that the DNA was extracted from a bone of a corpse inhumed for over 20 years.

\subsection{Weight of evidence in disputed paternity}

In this case of disputed paternity we want to compare the hypotheses:
\begin{center}
    \begin{tabular}{ll}
       $\Hp$: The father
of $A$ has contributed to the mixture {\emph{vs.}}\\
      $\H_0$: no contributor to the mixture is related to $A$.    
   \end{tabular}
  \end{center} 
		
The evidence consists of $E = \{\mbox{\emph{cgt}}, \mbox{mixture}\}$,
where \emph{cgt} is the genotype of the alleged son $A$, and  the mixture consists of the  alleles and corresponding peak heights on all markers obtained from the EPG.

The weight of the evidence is reported as a likelihood ratio $\LR$
\begin{equation}
	 \LR=\frac{L(\Hp)}{L(\H_0)}=\frac{\Pr( E \cd \Hp)}{\Pr(E \cd \H_0)}.
	\label{eq:lr}
\end{equation}
The likelihood ratio $\LR$, termed the paternity index, was introduced by
 \textcite{Essen-M}, who also gave a guideline transforming  the $\LR$ and posterior probability, based on uniform priors, onto a scale of verbal predicates. 
 
With uniform prior probabilities $\Pr(\Hp)=\Pr(\H_0)$, we have by Bayes's theorem the posterior probability of paternity
\begin{equation}
\Pr(\Hp|E) = \frac{ \LR} {(1+ \LR)}.
\label{eq:posterior}
\end{equation} 
\textcite{Essen-M} suggested a threshold of $0.9973$ (a \LR\ of 370) for ``paternity practically proven'' when putative father, mother and child's DNA are available. In some  European countries, legislation sets a threshold (in Germany it is 0.999).  We prefer not to set a threshold probability but simply report the $\LR$.

\section{Methods for inference about relationships from DNA mixtures}
\label{sec:methods}

Let $U_i=U$ be a specified contributor to the mixture, and let \Ugt\ denote the genotype of $U$.
We are interested in assessing a potential relationship between $U$ and one or more other individuals \emph{who have a known relationship to each other}; the genotype information on these other individuals is denoted $R$.

\figref{DAG} shows a directed acyclic graph (DAG), a pictorial representation for establishing a potential relationship with a contributor to the mixture. The blue nodes represent the gamma model for the peak heights. Specifically, node ($\phi, \rho, \xi, \eta$) corresponds to the model parameters, node $z$ to the peak heights and  $\mathbf{n}$ to vectors of allele counts representing all possible
combinations of genotypes, which in turn determine the distribution of the putative relatives' genotypes \textit{R}, under the hypothesis $\H$. For example, under the paternity hypothesis   $\Hp$, the putative father with genotype distribution \Ugt\ (green node) is the father of the alleged child with known genotype $\mbox{\emph{cgt}}$, a component of $R$. 

  We have that $R$ is conditionally independent of $\mathbf{z}$ given \Ugt\
$$
\CId{R}{\mathbf{z}}{\Ugt},
$$
 as implied by the DAG in \figref{DAG}. Two common examples are where $R$ denotes (i) the genotype of a child, or (ii) the genotypes of a child and its mother, where in both cases the potential relationship under test is that $U$ is the father of the child.

The hypothesis that $U$ does have the specified relationship with the individuals whose genotypes are in $R$ is $\Hp$; the contrary hypothesis   $\H_0$ is that $U$ is unrelated to the individuals whose genotypes are in $R$. We let 
$$
\LR_\Ugt = \frac{P(R|\Hp,\Ugt)}{P(R|\H_0,\Ugt)} = \frac{P(R|\Hp,\Ugt)}{P(R|\H_0)}
$$
since under $\H_0$, the individual $U_i$ is unrelated to those typed in $R$, so \Ugt\ and $R$ are independent.

Then our required  \LR\ (for $\Hp$ against $\H_0$) is
\begin{align}
\LR =& \frac{P(R,z|\Hp)}{P(R,z|\H_0)} = \frac{P(R,z|\Hp)}{P(R|\H_0)P(z|\H_0)} = 
\frac{\sum_\Ugt P(R,z|\Hp,\Ugt) P(\Ugt|\Hp)}{P(R|\H_0)P(z|\H_0)} \notag \\
=& \frac{\sum_\Ugt P(R|\Hp,\Ugt) P(z|\Hp,\Ugt) P(\Ugt|\Hp)}{P(R|\H_0)P(z|\H_0)} \notag \\
=& \frac{\sum_\Ugt P(R|\Hp,\Ugt) P(z|\Ugt) P(\Ugt)}{P(R|\H_0)P(z)} \notag \\
=& \sum_\Ugt \LR_\Ugt \times P(\Ugt|z). \label{wlr}
\end{align}
Conditional on the values of parameters $\phi, \rho, \xi, \eta$, the markers are independent, so the overall likelihood ratio is the product of (\ref{wlr}) over the markers.

In order to compute the likelihood ratio, for each marker, we present four different methods. These all address the same question, but strike different balances between structural and algebraic computation.

A first method, termed weighted likelihood ratio (WLR), uses the distribution of a contributor's genotype obtained from the mixture deconvolution and then computes the likelihood ratio algebraically.  
A second method, termed additional likelihood nodes (ALN) is a modification of the \textcite{cowell:etal:15} model, incorporating one or more additional auxiliary variables based on the relationship under question. 
The WLR and ALN methods are alternative computational approaches to calculating the required  \LR\ by first conditioning on \Ugt\ and then integrating out over the distribution of \Ugt\ given the peak height data, as in (\ref{wlr}). 

A third method, meiosis Bayesian Network (MBN), modifies the genotype Bayesian network by directly introducing meiosis or segregation indicators \cite{Thompson:00,Lauritzen:03}, maintaining the Markovian allele count representation. Thus, instead of computing $\LR_\Ugt$ algebraically, the Bayesian network is extended to include all the individuals described by $R$, and the evidence $R$ incorporated by explicitly setting the genotypes in this network.

The fourth method, replacing probability tables (RPT), modifies the relative probability tables based on the potential relationship we wish to establish: $\LR_\Ugt= P(R|\Hp,\Ugt)/P(R|\H_0) $ is inverted with the aid of Bayes theorem to give $P(\Ugt|\Hp,R)$, and these values used to replace the default $P(\Ugt)$ (based on Hardy-Weinberg equilibrium in the assumed population) in the network. 

Each of these approaches is elaborated in more detail below, for the specific example of paternity testing.

The last three methods give exact solutions. However, the first method yields a very good approximation, its accuracy limited only by the fact that the high-probability genotypes are identified in the   \Hugin\ deconvolution    code using simulation. 
In the  weighted likelihood ratio method the DAG of \figref{DAG}  is computed in two separate parts; the blue and green nodes from the DNA mixture model; and the yellow nodes  for computing the likelihood ratio for paternity. The other methods use the entire DAG by introducing specific modifications to the Bayesian networks. 

\textcite{Kaur15} have presented a method to handle a paternity relationship based on DNA mixtures. Their method consists of enumerating the possible combination of genotypes in the mixture and then computing the likelihood ratio using a formula similar to (\ref{wlr}) but weighing the different potential genotypes by the allele frequencies in a database (rather than, as we have done correctly, by the posterior probabilities of genotypes given the EPG data).  \textcite{Chung2} describes familial searching on mixtures based solely on detected alleles and  not  accounting for artefacts. \textcite{Slooten:16} only consider sets of detected alleles for familial searching on DNA mixtures with dropout. The commercial software STRmix is claimed to implement familial searches against a database, for close relatives of contributors to mixed DNA.
Testing whether mixtures have a donor in common or whether a relative of one mixture donor is a donor to a second mixture has been investigated in \textcite{Ryan:etal:16} and  \textcite{Slooten:17}. 
 All these papers deal with simpler scenarios to those analysed here; few of them use peak height information at all, and none of those that do incorporate it fully coherently through a probabilistic model.

\subsection{WLR method}
In the WLR method, the required distribution for \Ugt\ given the mixture data is obtained by deconvolution, leading to an approximation to $P(\Ugt|z)$. The \Ugt-specific likelihood ratios $\LR_\Ugt$ are derived algebraically, and the weighted sum (\ref{wlr})
computed.

\subsection{ALN method}
In the ALN method, an additional likelihood node is introduced into the BN, the allele counts for the specified contributor to the mixture as its parents. The values of $\LR_\Ugt$ are used in defining the CPTs for this node, as described in more detail below, and the weighted sum (\ref{wlr}) then implicitly computed during the equilibration of the network.

\subsection{Example: mother and child genotyped}\label{sec:cgtandmgt}
Here the relationship data $R$ represents the genotypes of two individuals, a child and its known mother. Under $\Hp$, the father is $U_i$, while under $\H_0$ the father is an unknown random member of the population. As usual, in deriving the \Ugt-specific likelihood ratios $\LR_\Ugt$ we can work marker-by-marker. 

In the absence of mutation, we are only concerned with cases where the mother and child genotypes have alleles in common. Then all relevant combinations of \cgt\ and \mgt\ are covered by the following
$$
\LR_\Ugt = \frac{P(\cgt|\mgt,\Ugt,\Hp)}{P(\cgt|\mgt,\H_0)}
=\begin{cases}
n_{ia}/2q_a& \text{if } \cgt=\{a,a\}, \mgt=\{a,a\} \text{ or } \{a,b\} \\
n_{ib}/2q_b & \text{if } \cgt=\{a,b\}, \mgt=\{a,a\} \text{ or } \{a,c\} \\
(n_{ia}+n_{ib})/(2(q_a+q_b)) & \text{if } \cgt=\mgt=\{a,b\} 
\end{cases}
$$
where $a,b,c$ are distinct alleles and $q_a$ and $q_b$ are the  allele frequencies in the population. Note that the only allele counts for $U_i$ that appear explicitly as parents when these expressions are used in defining the CPT for the likelihood node are those ($a$ and/or $b$) in \cgt.

\subsection{Example: only child genotyped}\label{sec:onlycgt}
If the mother's genotype is not available,  $R$ represents only the  genotype of the child. The hypotheses $\Hp$ and $\H_0$ are as before. 

By similar logic, we find
\begin{equation}\label{eq:lrugt}
\LR_\Ugt = \frac{P(\cgt|\Ugt,\Hp)}{P(\cgt|\H_0)}
=\begin{cases}
n_{ia}/2q_a & \text{if } \cgt=\{a,a\} \\
n_{ia}/4q_a+n_{ib}/4q_b & \text{if } \cgt=\{a,b\}
\end{cases}
\end{equation}
where $a,b$ are distinct alleles. Again, the only allele counts for $U_i$ that appear explicitly as parents when these expressions are used in defining the CPT for the likelihood node are those ($a$ and/or $b$) in \cgt.

\begin{figure}[htbp]
\begin{center}
       \resizebox{0.7\textwidth}{!}{\includegraphics{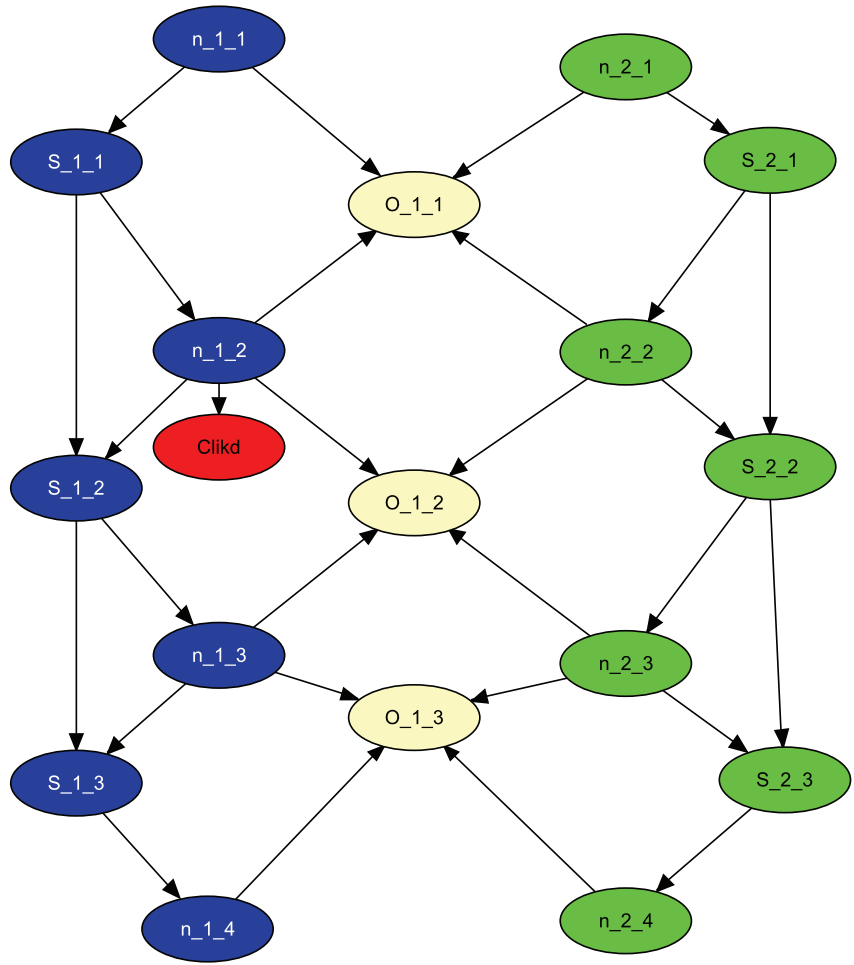}}
\end{center}
\caption{Bayesian network representation for an application of the ALN method, showing the additional child likelihood node for a case where there are $A=4$ alleles, and $\cgt=\{2,2\}$.}
\label{fig:method4}
\end{figure}

\figref{method4} shows a Bayesian network representation, as in \textcite{cowell:etal:15}, under paternity $\Hp$ for  a homozygous child with genotype $\cgt=\{2,2\}$,   but with an additional likelihood node \textit{Clikd} (red). This is a graphical child of the father's relevant allele counts. For a heterozygous child $\cgt=\{a,b\}$, the \textit{Clikd}  node is graphical child of the father's allele counts $n_a$ and $n_b$ and a boolean node, which switches between the two allele values. The  network is constructed  after knowing the child genotype, so \textit{Clikd} is linked to at most two allele count nodes, thus avoiding creating big cliques.

The blue/green nodes  in \figref{method4} refer to the Markov genotype representation of the first/second contributor to the mixture via the allele counts $n_{ia}$ and their partial sums $S_{ia}$. The auxiliary boolean $O_{ia}$ nodes allow the exact evaluation of the likelihood function $L_{ma} (z_{ma})$ in (\ref{eq:dens}) for the peak heights by probability propagation. 
For further details see \textcite{cowell:etal:15}.

\subsection{MBN method}
\label{sec:meiosis}

\begin{figure}[htbp]
\begin{center}
       \resizebox{\textwidth}{!}{\includegraphics{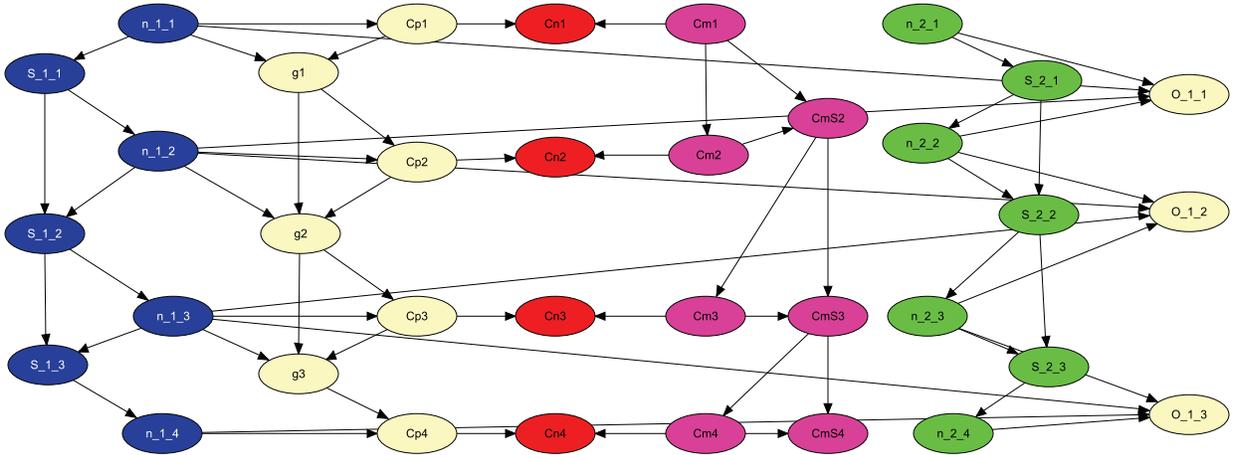}}
\end{center}
\caption{Meiosis Bayesian network representation of father--child relationship for a case with $A=4$ alleles, showing that is  Markovian over alleles.}
\label{fig:method2}
\end{figure}

\figref{method2} shows the meiosis Bayesian network representation for a single marker, under the paternity hypotheses $\Hp$ for a subset  $A=4$ alleles. The network  is Markovian over allele values.  The blue nodes refer to the father's allele counts  $n_{1a}$ and their cumulative sums $S_{1a}$.  The pink nodes refer to the child's maternal allele counts $Cm_a$ and their cumulative sums $CmS_a$, so unlike the other allele counts these sum to 1   over alleles (not 2 as other cumulative sums).  The red nodes are  the alleged son's allele counts $Cn_{a}$. They are simply sums of maternal $Cm_a$  and paternal  $Cp_a$ allele counts.   
The novelty is in the way that meiosis is captured for the child's paternal allele counts
 using the $g$ nodes.

Each node $g_a$  takes values 0, 1, 2, where 
\begin{itemize}
	\item $g_a= 0$  means that one of the alleles $1, 2, \ldots,a$ is present in the father, and that this
allele has been passed to the child.
\item $g_a= 1$ means that none of the alleles $1, 2, \ldots,a$ is present in the father.
\item $g_a= 2$ means that one of the alleles $1, 2, \ldots,a$ is present in the father, and that this
allele has not been passed to the child.
\end{itemize}

The novel conditional probability tables are defined by:
$$P(Cp_a = 1 \cd n_{1a}, g_{a-1}) = \left\{\begin{array}{cr} 0 \; \; &\mbox{if} \; \; n_{1a} = 0\\
                                             g_{a-1}/2 \; \; &\mbox{if} \; \; n_{1a} = 1\\
                                             1 \; \; &\mbox{if} \; \; n_{1a} = 2 \end{array}\right.$$
of course, $\Pr(Cp_a = 0 \cd n_{1a}, g_{a-1})  = 1- \Pr(Cp_a = 1 \cd n_{1a}, g_{a-1})$, while  
$\Pr(g_a \cd  n_{1a}, Cp_a , g_{a-1})$, is
defined by the deterministic relationship:
$$
g_a \cd n_{1a},  Cp_a , g_{a-1} =\left\{\begin{array}{cr} 
2 &\; \;\mbox{if} \; \; n_{1a}\ge 1, Cp_a =0 \; \; \mbox{and} \; \; g_{a-1}=1\\
0 &\; \;\mbox{if} \; \;n_{1a}\ge 1, Cp_a =1 \; \;\mbox{and} \; \; g_{a-1}=1\\
g_{a-1} & \; \;\mbox{otherwise.} \end{array}\right.$$

The second contributor to the mixture's allele counts $n_{2 \bullet}$ and their partial sums $S_{2 \bullet}$  are shown as green nodes. Finally, the boolean nodes $O_{1 \bullet}$ are where the information about the peak heights is incorporated.

\subsection{RPT method}
\label{RPT}
This method replaces the default $P(\Ugt)$ tables (based on Hardy-Weinberg equilibrium in the assumed population) in the network,  with tables for $P(\Ugt|\Hp,R)$,  e.g.  the father's genotype tables, given $\cgt$. The Markovian genotype structure is maintained.

The $S_{ia}$ are re-defined to be the cumulative sums of the $n_{ia}$ excluding the IBD allele (the one passed to the child), so its values can only be 0 and 1. 

In the homozygous case, suppose the child has genotype $(a',a')$. Then the father must be $(a',a)$ where $a$ is drawn from the gene pool. The binomial distribution given in equation (2.4.1) of \textcite{cowell:etal:15} is replaced by

$$n_{i,a+1}|S_{ia} \sim \delta_{a+1,a'}+\mathrm{Bin}\left(1-S_{ia}, q_{a+1}/\sum_{b > a}q_{b}\right)$$
where 
$$
{\delta_{i,j}} = \left\{\begin{array}{cr} 0 \; \; &\mbox{if} \; \; i \ne j\\
                                            1 \; \; &\mbox{if} \; \; i = j \end{array}\right.
$$
For $a=a'$, the table for  $S_{ia}$ is redefined to suit the new definition of $n_{ia}$, for the other values of a, the existing tables (created by functions in \DNAmixtures) are correct already.

In the heterozygous case, the child is say $(a',b'), a' \neq b'$,  an extra boolean node is introduced, `\ibdyet', with no parents and probabilities $(.5, .5)$. The children of this node are $n_{ia}$ and $S_{ia}$ for $a = a'$ and $b'$. The role of this node is to discriminate between the cases where it is $a'$ or $b'$ that the father has passed to the child. If \ibdyet\ is True then
$$n_{i,a+1}|S_{ia} \sim \delta_{a+1,a'}+\mathrm{Bin}\left(1-S_{ia}, q_{a+1}/\sum_{b > a}q_{b}\right)$$
while if it is False
$$n_{i,a+1}|S_{ia} \sim \delta_{a+1,b'}+\mathrm{Bin}\left(1-S_{ia}, q_{a+1}/\sum_{b > a}q_{b}\right)$$
and the tables for $S_{ia'}$ and $S_{ib'}$ are modified accordingly. 

\section{Results for alleged father in mixture}
\label{sec:paternity}

\subsection{Child only typed}
\label{sec:results1}
In this section we demonstrate the results and performance of our methods on the case study presented in Section \ref{sec:example}, based on the complete data on 17 markers (including Amelogenin) in the NGM SElect{\textsuperscript{TM}} PCR Amplification kit.  Here we assume  known allele frequencies from the
Italian population  \cite{NGM_freq2013,ID_freq2006}, and adopt a threshold of $C=0.001$ rfu.\footnote{\textcite{mortera:16} used a higher threshold.} 

Unless otherwise stated all computations are made conditional on the information that the major contributor $U_1$ to the DNA mixture is a male. In the case, as here, where the AMEL marker is among those included in the mixture, the evidence that the putative father $U_i$ is Male is introduced by setting the allele count nodes $n_{i,a}=1$ for each of the alleles $a=X$ and $Y$, in the BN, in addition to the other modifications to the BN used in most of our 4 methods.

The MLEs of the parameters that characterize the DNA mixture model, together with their approximate standard errors are given in \tabref{estimates}.  Here we assume that there are  two unknown contributors, $U_1$ and $U_2$, to the DNA mixture. The results on relationship inference presented below use parameters fixed at the MLE values in \tabref{estimates}. 

The estimated proportion $\phi_{U_1}$ of DNA contributed to the mixture by the major contributor $U_1$ is roughly 98\%.

\begin{table}[htbp]
\caption{Parameter estimates and approximate standard errors for 2 unknown contributors with maleness evidence, for the example in Section \ref{sec:results1}.}
\label{tab:estimates}
\begin{center}
  \begin{tabular}{lrr}
Par.&Est.&SE\\
\hline
$\mu$ &    807  &  163\\
$\sigma$   &   1.18 & 0.14\\
$\xi$     &  0.007 &0.006\\
$\phi_{U_1}$ & {0.978}  & 0.013\\
$\phi_{U_2}$& 0.022 &  0.013\\
\hline
\end{tabular}
\end{center}
\end{table}

\begin{table}[htbp]
\caption{Extract of the top-ranking genotypes and their probability, for the example in Section \ref{sec:results1}.}
\label{tab:top}
 \begin{center}
  \begin{tabular}{l|rr|rr|c}
	    Marker &	\multicolumn{2}{c|} {Top-ranked}	 & \multicolumn{2}{c|}{Alleged son's} &  {Probabilities without} \\
	     &   \multicolumn{2}{c|} {genotype}       &  \multicolumn{2}{c|} {genotype, \emph{cgt}} &  
			and with maleness\\
			\hline
D16S539   &	 {11}  &	12	& 10	  & {11}&  0.9669, 0.9671 \\
D8S1179    &	13      &	 {15} &	 {15}&	16 & 0.7279, 0.7292 \\
D21S11   &	 {30} &	31.2 & {30} & 32  & 0.3528, 0.3531\\
D18S51   &	12	& {13}	  & {13}	&16 & 0.9815, 0.9816 \\
$\cdots$& $\cdots$ & $\cdots$ &$\cdots$&$\cdots$ & $\cdots$, $\cdots$\\
SE33 &	 {20}	&  {20}  &	14 & {20} & 0.9925, 0.9926\\
\hline
\end{tabular}
\end{center}
\end{table}

We first illustrate the WLR method applied to the paternity case. \tabref{top} shows that $U_1$'s top-ranking  predicted genotype is compatible with \cgt\ on all markers. The additional information on the maleness slightly increases the probability of the top-ranked genotype. All the predictive probabilities are greater than 0.5 except for marker D21S11.

\tabref{D21S11} shows the ranking of $\Ugt$ with corresponding predictive probability for a  marker D21S11. The first three genotypes are compatible with \cgt\, whereas from rank 4 on they are not, yielding a null contribution to the $\LR$.

\begin{table}[htbp]
\caption{Ranking of $\Ugt$  with corresponding predictive probability for a  marker D21S11, for the example in Section \ref{sec:results1}.
}
\label{tab:D21S11}
\begin{center}
  \begin{tabular}{l|l|ll|r|r|r}
Marker & {Rank} &	\multicolumn{2}{c|}{$\Ugt$}	&	Prob. & $\Pr(\mbox{\emph{cgt}} \cd \Ugt , \Hp)$& $\Pr(\mbox{\emph{cgt}} \cd  \H_0)$ \\
	   			\hline		
 D21S11 & 1 &  {30} & 31.2 & 0.353 & 0.0055 & 0.0051 \\
     & 2 &   {30} & 34& 0.258 & 0.0055 & 0.0051 \\
     & 3 &  {30} & 31 & 0.190 & 0.0055 & 0.0051 \\
     & 4&  31 & 31.2 & 0.079 & 0 & 0.0051 \\
     &5 &  31 & 34   & 0.058 & 0 & 0.0051 \\
    &6 & 29 & 31 & 0.053 & 0  &  0.0051 \\
     & 7 & 34 &31.2 &0.0047 & 0 & 0.0051 \\
    & 8 &  29 & 31.2 & 0.0022 & 0 & 0.0051 \\
    & 9 &  29 & 34    & 0.0016 &  0 & 0.0051 \\
\hline
\end{tabular}
\end{center}
\end{table}


\tabref{comparison} shows a comparison between the top-ranked genotypes from the deconvolution, and the methods WLR, ALN, MBN and RPT.
\begin{table}[htbp]
\caption{Comparison between marker-wise likelihood ratios and overall $\LR$ for top-ranked genotypes and methods WLR, ALN, MBN and RPT, for the example in Section \ref{sec:results1}.}
\label{tab:comparison}
 \begin{center}
  \begin{tabular}{l|rr|rr|r|r|r}
	      & & & & &\multicolumn{3}{c} {Likelihood ratios} \\
Marker &	\multicolumn{2}{c|} {Top-ranked}	 & \multicolumn{2}{c|}{Alleged } & Top-ranked & WLR & ALN MBN \\
	     &   \multicolumn{2}{c|} {$\Ugt$}       &  \multicolumn{2}{c|} { $\cgt$} &  &  &  \& RPT\\
			\hline
D16S539   &	 {11}  &	12	& 10	  & {11}& 0.761  & 0.744 & 0.744 \\
D8S1179    &	13      &	 {15} &	 {15}&	16 & 1.76  &  1.51 & 1.51\\
D21S11   &	 {30} &	31.2 & {30} & 32  & 1.08 & 0.869 & 0.869 \\
D18S51   &	12	& {13}	  & {13}	&16 & 1.70 & 1.72 & 1.72\\
$\cdots$& $\cdots$ & $\cdots$ &$\cdots$&$\cdots$& $\cdots$ & $\cdots$ & $\cdots$\\
SE33 &	 {20}	&  {20}  &	14 & {20} & 10.18 & 10.14 & 10.14 \\
\hline
$\log_{10} \LR$ &  &    &     &        & 5.6708 & 5.4253 & 5.4251 \\
\hline
\end{tabular}
\end{center}
\end{table}
Using only the top-ranked $\LR$ does not take into account any uncertainty.
In this example, WLR is a good approximation to ALN, MBN and RPT \emph{which give exact results}.
In all cases the evidence in favour of the hypothesis of paternity is overwhelming. Under a uniform prior probability this would lead to a posterior probability of paternity of 0.999996.
 For any prior on $\H_p$ greater than 0.01 the posterior probability of paternity is greater than 0.9996,   extremely strong    evidence in favour of paternity.

\subsection{Mother typed too}
\label{sec:results1m}

For an illustration both of how genotype information on additional relatives can strengthen inference, and of the flexibility of our general approach, we augment our motivating disputed paternity example with a fictional genotype profile for the mother shown in the second and third columns of \tabref{comparison2}.  Here we do not consider the possibility of mutation.

\tabref{comparison2} shows a comparison between methods WLR, ALN, MBN and RPT, without and with information on  \mgt. The results for the WLR method with \mgt\ differ from the exact methods only in the 4th significant digit and are thus not given. The information on \mgt\ increases the overall $\LR$ roughly 540 times. 

The top-ranked profile for the father is in this case identical to that without the \mgt\ information, and if this profile were directly observed, the $\log_{10}$\LR\ for paternity would be 8.4022.
\begin{table}[htbp]
\caption{Comparison between marker-wise likelihood ratios and overall $\LR$ for the exact methods, ALN, MBN and RPT, with and without \mgt, for the example in Sections \ref{sec:results1} and \ref{sec:results1m}.}
\label{tab:comparison2}
 \begin{center}
  \begin{tabular}{l|rr|rr}
	 &\multicolumn{2}{c|} {Mother's genotype}      &\multicolumn{2}{c} {Likelihood ratios} \\
	Marker			  &  \multicolumn{2}{c|} { \mgt}  &\multicolumn{1}{c} {without \mgt}     &  \multicolumn{1}{c}{with \mgt}    \\
	     			\hline
D16S539  & 10 & 11 &	 0.744 & 1.25 \\
D8S1179  & 10 & 16   & 1.51 & 3.02 \\
D21S11 & 26 & 32   & 0.869 & 1.74 \\
D18S51 & 13 & 16  & 1.72 & 1.85 \\
$\cdots$& $\cdots$ & $\cdots$ & $\cdots$ & $\cdots$ \\
SE33 & 14 & 22 & 10.14 & 20.28\\
\hline
$\log_{10} \LR$  & & & 5.4251 & 8.1571 \\
\hline
\end{tabular}
\end{center}
\end{table}

\subsection{Computation time}

\tabref{times} gives a comparison among  the computation times,   listed in increasing order for    the 4 methods, for the task described in Section \ref{sec:results1}.   These were obtained on an Intel i7-4790 processor clocked at 3.60GHz. Here ALN runs the fastest, closely followed by RPT. In general, however, comparison between the methods will depend on the complexity of the relationship in question. For example, using the ALN method in a paternity case, the additional likelihood node is linked to only 1 or 2 allele counts. In more complex relationships one could need more links and computation would be slower. See also the Discussion, Section \ref{sec:discn}.

\begin{table}[htbp]
\caption{Comparison among computation times for the 4 methods, for the example in Section \ref{sec:results1}.}
\label{tab:times}
\begin{center}
\begin{tabular}{l|r}
 Method &	Time (seconds)	 \\
	  			\hline
ALN  &	1.32  \\
RPT  &  1.66   \\
MBN  &  2.82\\
WLR  & 46.90 \\
\hline
\end{tabular}
\end{center}
\end{table}

\section{Results for unknown in mixture, potential mother typed}
\label{sec:results2}

We have also analysed a criminal case where we have data on 3 crime traces, denoted $T_1, T_2$ and $T_3$,  amplified with the NGM amplification kit consisting of 17 markers including Amelogenin. We used US Caucasian allele frequencies \cite{butler:etal:03}. The genotype of the victim $V$ and \mgt, that of the alleged mother of a contributor to the mixture, were also available. We assume that there are at most 3 contributors to each mixture, the victim $V$ and two unknown contributors denoted by $U_1$ and $U_2$, with the unknown contributors labelled in the same way in each of the 3 traces.  Here we set the threshold to $C=50$ rfu. 

In this problem, the roles of child and parent are reversed compared to the situation of Section \ref{sec:onlycgt}: this time it is the child who is hypothesised to contribute to the mixture while the parent (the mother) is separately genotyped. However, under Hardy-Weinberg equilibrium, ignoring mutation, and in the absence of genotype information for the other parent, the genotypes of the child and mother (or father) are exchangeable random variables. So under both $\Hp$ and $\H_0$, the conditional distribution of the mother's genotype given that of the unknown contributor who may be her son, is as in (\ref{eq:lrugt}), with \cgt\ replaced by \mgt. Exactly the same codes can then be used to implement the ALN, RPT or WLR methods for the present problem.

\tabref{mle} shows the maximum likelihood estimates of the parameters based on the combined information from $T_1, T_2, T_3$. Note that in the first trace $T_1$ the major unknown contributor is estimated to have a fraction  $\phi_{U_1}=0.712$ of DNA,	more than  3 times that of  the victim  $\phi_{V}=0.221$, whereas the proportions of DNA they contribute to the second mixture $T_2$ are roughly equal, and in the third trace the victim contributes a greater amount  $\phi_{V}=0.626$  of DNA than $U_1$.

Furthermore, the second unknown contributor $U_2$, whose presence  can explain the presence of allelic dropin,
 contributes a small amount to  $T_1$ and $T_2$, but a negligible amount to the third trace $T_3$. The mean stutter proportion is estimated as around 13\% for $T_1$, reducing to less than half (4.8\%) in $T_2$ and almost vanishing in $T_3$. 

\begin{table}
\caption{Maximum likelihood estimates for the mixture parameters based on combined information on  $T_1, T_2, T_3$, for the example in Section \ref{sec:results2}.}
\label{tab:mle}
\begin{center}
\begin{tabular}{lr|r|r}
Parameter&$T_1$&$T_2$&$T_3$\\ 
\hline
$\mu$	&3858	&		1289	&			1836		\\
$\sigma$	&	0.408		&	0.671	&		0.562			\\
$\xi$	&	0.127	&	 	0.048	&		0			\\
$\phi_{V}$	&	 0.221		&	0.526		&	0.626	\\
$\phi_{U_1}$	&	0.712		&	0.448		&	0.374		\\
$\phi_{U_2}$	&	0.067		&	0.026		&	0	\\
\hline
\end{tabular}
 \end{center}
\end{table}

\begin{table}
\caption{Extract of \mgt\ and  the predicted profiles of  the major unknown contributor $U_1$ based on combined information on  $T_1, T_2, T_3$  with and without information on the mother's genotype, for the example in Section \ref{sec:results2}.}
\label{tab:sep2} 
\begin{center}
\begin{tabular}{l|rr|rr|rr}
& & & & & \multicolumn{2}{c}{probability} \\
Marker&\multicolumn{2}{c|}{\mgt}&\multicolumn{2}{c|}{$\Ugt$} & with \mgt & without \mgt\\ 
\hline
D21S11&29 &29 &29  &30.2&1& 0.9919\\
      &   &   &28  &30.2& &0.0072 \\
      &   &   &30.2&30.2& &0.0004 \\
      &   &   & 32 &30.2& & 0.0004\\
			\hline
D22S1045&15&16&15&16&0.9865&0.9865\\
&&&16&16&0.0096 &0.0096\\
&&&15&15&0.0039 &	0.0039	\\	
\hline
D2S441&11&14&	14 &11.3&1& 0.9989\\
          &&&11.3&11.3& & 0.0011\\
\hline
TH01&6&9&	6&9&0.9989&0.9987\\
&&&6&6&0.0011&0.0012\\
\hline
\end{tabular}
\end{center}
\end{table}

In this criminal case we might want to compare the hypotheses: 
\begin{center}
    \begin{tabular}{ll}
      $\Hp$: $U_1$ is the child of \mgt\    {\emph{vs.}}
      $\H_{{0}}$: no unknown contributors are related to \mgt\,    
    \end{tabular}
  \end{center} 
where $U_1$ is the major unknown contributor to the mixture. The following  inferences about relationships  are based on parameter values fixed at the MLEs of \tabref{mle}.
Using  the ALN  or RPT methods  gives a likelihood ratio in favour of $\Hp$ of  $\log_{10}  \LR= 5.275 $ (or $\LR=188330.3$).
If instead we were to compare the hypothesis  $\Hp$: $U_2$ is the child of \mgt\ to $\H_{{0}}$ the likelihood ratio would be much smaller,
  $\log_{10}$\LR= 1.569  (\LR=37.05). 

\tabref{sep2} shows a comparison of an extract of the  predicted profiles of the  major unknown contributor $U_1$ based on the combined information in  $T_1, T_2, T_3$  when analysis is made with and without information on the mother's genotype. For most markers, as for 
 D2S11 and  D2S441,  using the information on  $U_1$'s mother's genotype yields sharper predictions, but all very similar to those based solely on the three traces.


\tabref{single} shows a comparison between the likelihood ratios obtained for comparing the hypotheses  $\Hp$ to $\H_{{0}}$ by analysing each single mixture trace separately and the likelihood ratio we obtained before,  based on the combined evidence from the 3 traces. 
Trace $T_1$, where the proportion contributed by $U_1$ is around 70\% ($\phi_{U_1}= 0.712$)	 is very informative  and the \LR\ is slightly  greater than the \LR\ based on the combined evidence. Whereas, using   trace $T_2$    yields a $\LR$ about 539 times smaller than that based solely on $T_1$, and  using   trace $T_3$   alone yields a \LR\ about  1135 times smaller than that based on $T_1$.

\begin{table}
\caption{Comparison of the likelihood ratios based on $T_1, T_2$ and  $T_3$ separately and the likelihood ratio based on combining the information from  $T_1, T_2, T_3$, for the example in Section \ref{sec:results2}.}
\label{tab:single}
\begin{center}
\begin{tabular}{l|rrr|r}
&\multicolumn{3}{c|}{separate traces}&\multicolumn{1}{c}{combined traces}\\
&$T_1$&$T_2$&$T_3$ & \multicolumn{1}{c}{$T_1\&T_2\&T_3$}\\ 
\hline
$\LR$	            & 192578  	&	357.24 		&		 169.65	& 188330	\\
$\log_{10} \LR$	 	&  5.28   	&	2.55		 &		2.23   & 5.28\\
\hline
\end{tabular}
 \end{center}
\end{table}

\section{Leaving the contributor unspecified, and likelihood ratios for unions of alternative hypotheses}
\label{sec:unions}
In modelling of DNA mixtures, the contributors have to be labelled to ensure all parameters are identifiable; the convention used in \DNAmixtures\ is for the contributors to be numbered from 1, in decreasing order of the estimated proportion they contribute to the mixture in the first trace. This labeling affects the specification of hypotheses about relationships with the contributors. In our numerical examples, we have chosen to interpret, for example, the hypothesis that `a contributor to the mixture is the father of the specified child' as `contributor $U_1$ is the father of the specified child'. Exactly the same method could be used to evaluate similar hypotheses referring to $U_2$, $U_3$, etc.
This is partly for reasons of practicality and convenience: to deal precisely with a paternity hypothesis about an unspecified contributor to the mixture requires a more complicated BN, with increased storage and time requirements. 

In some situations, it is perfectly appropriate to formulate hypotheses about relationships in terms of the major contribution $U_1$; this is the case in the examples in Section \ref{sec:paternity}, where we believe that we have a DNA mixture dominated by DNA from the bones of the deceased singer, subsequently contaminated.

In other situations, we may prefer to assess 
a relationship with an unspecified contributor; this is formally a case of evaluating an alternative hypothesis $\Hp$ that is the union of two or more corresponding hypotheses about specified contributors, $\Hp=\H_1\cup\H_2\cup\cdots$, where $\H_k$ is the hypothesis that the relationship in question is with the $k$th contributor. The problem of defining a \LR\ for $\Hp$ against $\H_0$, given the likelihood ratio $\LR_k$ for $\H_k$ against $\H_0$, $k=1,2,\ldots$ is a generic one.  

A possible generic solution is to take $\LR=\max \LR_k$; this is similar to standard practice in evaluating generalised likelihood ratios in regular parametric problems, but is probably unacceptable in judicial work. More satisfactory in the present context would be to follow the Bayesian interpretation of the likelihood ratio.
A simple application of Bayes' theorem gives
$$
\LR = \frac{\sum_k \LR_k P(\H_k)}{\sum_k P(\H_k)}=\sum_k \LR_k P(\H_k|\Hp)
$$
which, as the second form above makes clear, depends on the relative prior probabilities for the alternative hypotheses, but not their absolute probabilities. If specifying these relative prior probabilities is difficult or impossible in context, appropriate bounds could be reported. In a criminal trial, one might suggest reporting $\min_k \LR_k$, to avoid exaggerating how incriminating the evidence is; but this may be considered over-conservative, especially if an exclusion is obtained for some $k$, when $\min_k \LR_k=0$. For a civil court, and possibly in some criminal trials, one could report the range of \LR\ over a reasonable range of priors. 

Applying these ideas to the example in Section \ref{sec:results2}, the relevant contributor-specific likelihood ratios are $\LR_1=188330.3 $ and $\LR_2=37.05$. Then $\min_k \LR_k=37.05$, while if we used priors with $P(H_1)=P(H_2)$ we would report the arithmetic mean $(188330.3+37.05)/2=94183.67$.

\section{Software}
\label{sec:software}

Our calculations are performed in \R\ using a suite of functions called \KinMix, available online at \url{https://people.maths.bris.ac.uk/~mapjg/KinMix/}. These additional functions call functions in the \RHugin\ package to augment the capabilities of the \DNAmixtures\ package.

\section{Discussion}\label{sec:discn}
This paper gives the first coherent way to model inference about relationships from DNA mixtures. The methods can be readily extended to analyse different scenarios. A variety of simple problems are illustrated.  The code for the analyses presented is available as supplementary material.  We also show how the additional information of specific genotypes relative to the relationship under analysis greatly strengthens the resulting inference. 
The  analysis concerning mixtures  with hypotheses on familial relationships
could also be useful for identifying disaster victims. Here we have treated the allele frequencies as fixed and  known, however,  the analysis  could  be extended to include uncertainty in allele frequencies as  shown in \textcite{green:mortera:09}.

We have also not considered the possibility of mutation. Each of our methods can be extended to handle mutation. There are simple extensions to the expressions for $\LR_\Ugt$ in both Sections \ref{sec:cgtandmgt} and \ref{sec:onlycgt} for any mutation model; these can readily be used in the WLR method. The modifications can also be used in coding the ALN method, for which there will need to be a link to the child likelihood node from $n_{ia}$ for every allele $a$ from which mutation is possible to an observed allele in the child genotype. In particular, for a one-step mutation model, there will be at most 6 links (or at most 3 if the child is homozygous). Adapting the RPT or MBN methods requires more significant elaborations to the Bayes nets. In each case there are likely to be increased computational costs.

Although the computation time comparisons in Table \ref{tab:times} suggest rather clear-cut preferences between the 4 methods presented, we believe that each of the 4 may be useful in extending our methods to particular further cases of relationship inference based on DNA mixtures. Whenever an algebraic expression for $\LR_\Ugt$ can be obtained reliably with a modest amount of labour, it seems preferable to adapt the ALN method to the new problem, as this expression appears explicitly in the code where the conditional probability tables are established; it may be necessary to add additional nodes and edges to the Bayes net, and this requires some understanding and expertise of \RHugin. The WLR method uses the same algebraic expression for $\LR_\Ugt$ explicitly, and requires no Bayes net coding or computation at all beyond that already set up in \DNAmixtures, but of course incurs a heavier computational price; it will be useful in prototyping. For the RPT method, it is necessary to invert the expression for $\LR_\Ugt$ to obtain a new conditional probability table for $\Ugt$, but once that is done, this method has similar computational cost to ALN, while retaining exactly the graph topology of \DNAmixtures. Finally, if the pedigree connecting all of the actors is sufficiently complicated that reliable algebraic derivation of $\LR_\Ugt$ is problematic, then the MBN method may be preferable; it does however entail much more extensive modification to the Bayes net.

Our emphasis here is on methodology and the general approach, but some qualitatitive conclusions might be drawn from the numerical results. For the examples in Sections \ref{sec:results1} and \ref{sec:results1m}, we saw that $\log_{10}\LR$ values of 5.6708
and 8.4022, respectively, would have been obtained if the top-ranked genotype profile for the putative father had this profile been directly observable for this individual. Based on the mixture evidence instead, these values become 5.4251 and 8.1571 (in each case, reduced by a factor of about 1.75 in the \LR).
In both settings, these represent extremely modest reductions in the weight of evidence, an encouraging sign for the usefulness of this kind of analysis. 

\section*{Acknowledgements}
The authors would like to thank the Isaac Newton Institute for
Mathematical Sciences, Cambridge, for support and hospitality during the programme \textit{Probability and Statistics in Forensic Science} which was supported by EPSRC grant number EP/K032208/1.  
We  thank the referees for his insightful and  helpful comments. We also thank Marjan Sjerps, Jacob de Zoete  and  Therese Graversen for useful discussions.

\section*{Appendix: Bayes net computations for DNA mixtures}

In this appendix, aimed at the more technically-interested reader, we give more details on the modelling and computational strategy introduced in Section \ref{sec:relinf}.

All of our proposed methods modify the Bayes net formulation of the model of \textcite{cowell:etal:15} and then, as in that earlier paper, perform the necessary computations to deliver the required likelihood ratios, as laid out by \textcite{graversen:lauritzen:comp:13}, appropriately generalised. 

With set values for the parameters, and with the observed peak heights $\mathbf{z}$ entered as data (via auxiliary boolean nodes as described in \textcite{graversen:lauritzen:comp:13}), all nodes in the network equilibrate to represent the marginal distributions of the corresponding variables, conditional on the values of $\phi, \rho, \xi, \eta$ and $z$. These distributions can be usefully interrogated, and the network elaborated if necessary to facilitate the delivery of distributions of other variables of interest, such as $\Ugt$, the genotype of a specified unknown contributor. 

For example, with the parameters in the model estimated via maximum likelihood, the peak heights and corresponding alleles in the  DNA mixture  can be used to deconvolve the mixture in order to predict, for each contributor to the mixed profile and for each marker, a set of possible genotypes, together with their marginal predictive probabilities. 

The methods devised in this paper, and fully described in Section \ref{sec:methods}, make use of this deconvolution, and other distributions obtained from the equilibrated BN, to make inference on putative relationships involving contributors to the mixture.

\bibliographystyle{oupvar}
\bibliography{refs,dna}
\end{document}